\documentclass[aps,twocolumn,tightenlines,nofootinbib,superscriptaddress]{revtex4-2}
\usepackage[utf8]{inputenc}
\usepackage{graphicx}
\usepackage{subcaption}
\usepackage{mwe}
\usepackage{mathrsfs}
\usepackage{amsfonts}
\usepackage{setspace}
\usepackage{cellspace}
\usepackage{amsmath,amssymb,bm}
\usepackage[colorlinks=true,linkcolor=blue]{hyperref}
\usepackage{xcolor}
\usepackage{epsfig}
\usepackage{slashed}
\usepackage{caption}
\usepackage{hhline,multirow,tabularx}  
\usepackage{dcolumn}    
\usepackage{url}        
\usepackage{braket}     
\renewcommand{\bra}[1]{\left<#1\left|}
\renewcommand{\ket}[1]{\right|#1\right>}

\newcommand{\ui}[1]{^{(#1)}}

\DeclareMathAlphabet{\mathdutchcal}{U}{dutchcal}{m}{n}
\SetMathAlphabet{\mathdutchcal}{bold}{U}{dutchcal}{b}{n}
\DeclareMathAlphabet{\mathdutchbcal}{U}{dutchcal}{b}{n}

\begin{document}

\title{Proton's gluon GPDs at large skewness and gravitational form factors \\ from near threshold heavy quarkonium photo-production}

\author{Yuxun Guo}
\email{yuxunguo@umd.edu}
\affiliation{Maryland Center for Fundamental Physics,
Department of Physics, University of Maryland,
College Park, Maryland 20742, USA}
\affiliation{Nuclear Science Division, Lawrence Berkeley National
Laboratory, Berkeley, CA 94720, USA}

\author{Xiangdong Ji}
\email{xji@umd.edu}
\affiliation{Maryland Center for Fundamental Physics,
Department of Physics, University of Maryland,
College Park, Maryland 20742, USA}

\author{Feng Yuan}
\email{fyuan@lbl.gov}
\affiliation{Nuclear Science Division, Lawrence Berkeley National
Laboratory, Berkeley, CA 94720, USA}

\begin{abstract}
We study the exclusive near-threshold photo-production of heavy quarkonium in the framework of the generalized parton distribution (GPD) factorization, taking the $J/\psi$ production as an example. Due to the threshold kinematics, the Compton-like amplitudes 
are related to gluon GPDs at large skewness $\xi$, distinct from the common kinematics in asymptotic high energy where the skewness is typically small. We discuss the nature of large-$\xi$ expansion of these amplitudes in terms of the moments of gluon GPDs in the large-$\xi$ limit. Based on that, we propose several ways to extract the first few moments of the gluon GPDs from these amplitudes, with the leading ones corresponding to the gluonic gravitational or energy-momentum tensor form factors (GFFs). We apply these methods to analyze the recent near-threshold $J/\psi$ production measurements by the $J/\psi$ 007 experiment and GlueX collaboration, and find that the $\xi$-scaling of the measured differential cross sections is consistent with the asymptotic behavior. However, the current data are not accurate enough yet for a complete determination of the gluonic GFFs, and therefore we consider some prospects for better extractions in the future.
\end{abstract}

\maketitle

\section{Introduction}

In the past decades, it has been realized how important a role gluons play inside the nucleon with quantum chromodynamics (QCD). The exclusive productions of heavy quarkonium off the nucleon have therefore drawn rising attention for their accessibility to the gluonic structure in the nucleon. Extensive programs measuring the exclusive meson production have been planned with the future Electron-Ion Collider (EIC)~\cite{Accardi:2012qut,Boer:2011fh,AbdulKhalek:2021gbh}. 
Meanwhile, 
near-threshold heavy quarkonium photo-production, 
e.g., $J/\psi$ production, has been carried out 
recently by experiment groups at Jefferson Laboratory (JLab)~\cite{GlueX:2019mkq,Duran:2022xag,GlueX:2023pev}. 
Near-threshold productions provide unique opportunities to study the nucleon structure at large momentum transfer, distinct from the common deeply virtual measurements where momentum transfer is considered small at high energy in the collinear limit. This has spurred many theoretical developments aiming to extract the critical information on the gluonic structures from such processes~\cite{Voloshin:1978hc,Gottfried:1977gp,Appelquist:1978rt,Bhanot:1979vb,Kharzeev:1995ij,Kharzeev:1998bz,Guo:2021ibg,Guo:2023pqw,Kharzeev:2021qkd,Gryniuk:2016mpk,Gryniuk:2020mlh,Hatta:2018ina,Hatta:2019lxo,Mamo:2019mka,Mamo:2021krl,Mamo:2022eui,Sun:2021pyw,Sun:2021gmi}. In addition, other production mechanisms that include the coupled channel contributions and resonance pentaquark state contributions have also been considered in Refs.~\cite{Winney:2023oqc, Du:2020bqj}.

In this work, we focus on the generalized parton distribution (GPD) factorization framework~\cite{Guo:2021ibg}, where the near-threshold photo-productions of heavy quarkonium are expressed in terms of the gluon GPDs in the heavy quark limit. In this case, the large momentum transfer also indicates large skewness for the GPD, i.e., the momentum transfer will be mainly in the longitudinal direction near the threshold. It has been argued that in the limit $\xi\to 1$, the amplitude will be dominated by the leading moments of GPDs which correspond to the gravitational form factors (GFFs)~\cite{Guo:2021ibg}. However, realistically, one has $\xi_{\rm{th}}\sim 0.6$ at the threshold in the case of $J/\psi$ production. Therefore, more careful study examining such relations away from $\xi=1$ is in need, for which the general behavior of GPDs at large $\xi$ will be of interest.

The large-$\xi$ behavior of GPDs has not been discussed much in the literature as most GPD studies are within the collinear factorization framework with small momentum transfer squared $t$ and small skewness parameter $\xi$ accordingly. Partonic interpretation of GPDs will go through an important transition from the forward limit $\xi\to 0$ to the large-$\xi$ limit $\xi \to 1$~\cite{Ji:1998pc,Brodsky:2000xy}. When $|x|>\xi$, GPDs resemble the parton distribution functions (PDFs) which are interpreted as the amplitudes of emitting and reabsorbing a parton, whereas for $|x|<\xi$ they resemble the distribution amplitudes (DAs) and are interpreted as the amplitudes of emitting/absorbing a parton-antiparton pair~\cite{Ji:1998pc}. In the large-$\xi$ limit, GPDs will be dominated by the DA-like region $|x|<\xi$, which does not couple to the PDF-like region that we know well from the forward limit.

Despite the lack of knowledge about the large-$\xi$ behavior of GPDs, in this work we attempt to extract the important information on the gluonic structures from the Compton-like amplitude at large $\xi$ utilizing the endpoint constraints that GPDs must vanish at $|x|=1$. In the previous work~\cite{Guo:2021ibg,Duran:2022xag,Guo:2023pqw}, the analyses are based on the Taylor expansion of the Wilson coefficients:
\begin{equation}
    \frac{1}{x+\xi}- \frac{1}{x-\xi} = \frac{2}{\xi} \sum_{n=0}^\infty \left(\frac{x}{\xi}\right)^{2n}\ ,
\end{equation}
as also suggested for the strange quark in Ref. \cite{Hatta:2021can}. However, this Taylor series only converges for $x\le \xi$ which does not generally hold. Therefore, concerns about its applicability as well as the reliability of the extracted leading terms arise naturally, which could damage the very foundation of all the extractions based on it.

In this work, we argue that the Compton-like amplitudes at large $\xi$ can be considered as an asymptotic series in terms of the moments of GPDs. As we will show, in the $\xi\to 1^-$ limit, the real parts of the Compton-like amplitudes can be asymptotically written as,
\begin{align}
\begin{split}
    \text{Re}\mathcal{H}_{g\rm{C}}(\xi,t)&= \mathdutchcal{C}_g(t) + \xi^{-2} \mathdutchcal{A}_g\ui2(t)+\xi^{-4} \mathdutchcal{A}_g\ui4(t)+\cdots\ ,
\end{split}\\
\begin{split}
    \text{Re}\mathcal{E}_{g\rm{C}}(\xi,t)&= -\mathdutchcal{C}_g(t) + \xi^{-2} \mathdutchcal{B}_g\ui2(t)+\xi^{-4} \mathdutchcal{B}_g\ui4(t)+\cdots\ ,
\end{split}
\end{align}
whereas the imaginary parts vanish. Such behaviors predict unique features in the $\xi$-dependence of the amplitudes and the differential cross sections correspondingly. Consequently, in the context of asymptotic expansion, the previous analyses are justified in the large-$\xi$ region even though the series itself might diverge. Additionally, the knowledge of the $\xi$-dependence can be used to account for the higher-moment contributions in the extraction of gluonic GFFs, which are one of the most important higher-order corrections. 

As we will show, this $\xi$-dependence from asymptotic expansion is consistent with the recent near-threshold $J/\psi$ photo-production measurements by the $J/\psi$ 007 experiment~\cite{Duran:2022xag} and the GlueX collaboration~\cite{GlueX:2023pev}. Based on this, we consider several scenarios for the examination of this framework as well as the possible extraction of the gluonic GFFs with it. The previous leading-moment approximation corresponds to the simplest scenario here, which can be improved, especially when more $\xi$-dependence of the data will be obtained. 

The organization of the paper is as follows. In sec. \ref{sec:asymptotic}, we first briefly review the GPD framework for the near-threshold heavy quarkonium production, and then we study the large-$\xi$ behaviors of GPDs and the corresponding asymptotic expansion of the amplitude. In sec. \ref{sec:dataanaly}, we consider the two recently published measurements by the $J/\psi$ 007 experiment~\cite{Duran:2022xag} and the GlueX collaboration~\cite{GlueX:2023pev} and study their $\xi$-scaling behaviors. We propose several scenarios for examinations and the extraction of gluonic GFFs utilizing these observations. In sec. \ref{sec:currentstatus}, we summarize the current status of the gluonic GFF extraction in the GPD framework and comment on the impact of the future experimental developments. In the end, we conclude in sec. \ref{sec:conc}.

\section{Heavy quarkonium threshold production, large-\texorpdfstring{$\xi$}{XI} GPDs, and asymptotic expansion}

\label{sec:asymptotic}
We start with a brief review of the GPD framework for the near-threshold heavy quarkonium production presented in Ref. \cite{Guo:2021ibg} for completeness. It has been shown therein that near the threshold the differential cross section can be written as:
\begin{equation}
\label{eq:xsec}
\begin{split}
\frac{\text{d} \sigma}{\text{d} t}= \frac{ \alpha_{\rm EM}e_Q^2}{ 4\left(W^2-M_N^2\right)^2}\frac{ (16\pi\alpha_S)^2}{3M_V^3}|\psi_{\rm NR}(0)|^2 |G(\xi,t)|^2\ ,
\end{split}
\end{equation}
where the hadronic matrix element $G(\xi,t)$ can be expressed as the convolution of the gluon GPDs $F_g(x,\xi,t)$ and the Wilson coefficient ${\cal A}(x,\xi)$:
\begin{align}\label{eq:amplitude2}
G(\xi,t)=\frac{1}{2\xi} \int_{-1}^1\text{d}x{\cal A}(x,\xi)F_g(x,\xi,t) \ ,
\end{align}
where the leading-order Wilson coefficient reads,
\begin{equation}
   {\cal A}(x,\xi)\equiv\frac{1}{x+\xi-i0}- \frac{1}{x-\xi+i0}\ .
\end{equation}
The gluon GPD $F_g(x,\xi,t)$ is defined as
\begin{align}
\label{gluonGPD}
   &F_g(x,\xi,t) \equiv\nonumber \\ &\frac{1}{(\bar P^+)^2} \int \frac{\text{d}\lambda}{2\pi} e^{i\lambda x}\bra{P'}F^{a+i}_{\;\;\;\;\;}\left(-\frac{\lambda n}{2}\right)F^{a+}_{\;\;i}\left(\frac{\lambda  n}{2}\right) \ket{P}\ ,
\end{align}
which can be parameterized as~\cite{Ji:1996ek,Diehl:2003ny}
\begin{align}
    F_{g}(x,\xi,t)=\frac{1}{2\bar P^+}\bar u(P')\left[H_g \gamma^+ +E_g\frac{i\sigma^{+\alpha}\Delta_{\alpha}}{2 M_N}\right]u(P) \ ,
\end{align}
where $H_g$ and $E_g$ are the well-known $H_g(x,\xi,t)$ and $E_g(x,\xi,t)$ GPDs.

In the heavy quark limit with $M_V\to\infty$, the momentum transfer squared $|t|$ approaches infinity near the threshold and $\xi\to 1$ accordingly. Then the $G(\xi,t)$ can be expanded as a power series of $x/\xi$, leading to~\cite{Hatta:2021can,Guo:2021ibg}
\begin{align}\label{eq:moments}
    G(\xi,t)=\sum_{n=0}^{\infty}\frac{1}{\xi^{2n+2}} \int_{-1}^{1} \text{d}x x^{2n}F_g(x,\xi,t) \ ,
\end{align}
which converges for $\xi\ge 1$. Under the approximation that the series is dominated by the leading terms, such processes serve as the probe of the gluonic GFFs. However, since the summation in eq.(\ref{eq:moments}) could be divergent in the realistic case where $\xi<1$, it should be understood as a formal summation rather than an actual one in the most general case. In this sense, the left-hand side will be the analytical continuation of the summation on the right-hand side when it does not converge. Consequently, the series itself may be regarded as an asymptotic series that does not generally converge. 

To examine this, note that eq. (\ref{eq:moments}) can be rewritten with the Mellin moments of the gluon GPD $F_g(x,\xi,t)$ defined as,
\begin{align}\label{eq:MelMom}
    \mathcal{F}_g\ui{n}(\xi,t)\equiv  \int_{0}^{1} \text{d}x x^{n-1}F_g(x,\xi,t) \ ,
\end{align}
to be
\begin{align}\label{eq:moments2}
    G(\xi,t)=\sum_{n=0}^{\infty}\frac{2}{\xi^{2n+2}} \mathcal{F}_g\ui{2n+1}(\xi,t) \ .
\end{align}
One sufficient but not necessary condition for it to converge is that $\mathcal{F}_g\ui{2n+1}(\xi,t) < \xi^2 \mathcal{F}_g\ui{2n-1}(\xi,t)$ as $n\to \infty$. To illustrate the higher-moment behavior, we start by considering the near-forward region of the GPD with small/medium $\xi$ and fixed $t$. Omitting the $\xi$- and $|t|$-dependence in this case,\footnote{According to the polynomiality condition~\cite{Ji:1998pc}, these Mellin moments $\mathcal{F}_g\ui{n}(\xi,t)$ depend on $\xi$ non-trivially, i.e., they are finite-order polynomials of $\xi$. However, for small/medium $\xi$, they are dominated by the lowest order in $\xi$, namely the $\xi^0$ terms, while the other terms are suppressed by powers of $\xi$. } the gluon GPD $F_g(x,\xi,t)$ near the endpoint $x=1$ can be written as,
\begin{align}\label{eq:asympx1}
    F_g(x,\xi,t) \sim (1-x)^\beta \text{~~as~~} x\to 1\ ,
\end{align}
with $\beta>0$ according the requirement that GPDs vanish at endpoints. Correspondingly, the Mellin moments will behave as,
\begin{align}\label{eq:asympmoment}
    \mathcal{F}_g\ui{n}(\xi,t) \sim n^{-\beta-1} \text{~~as~~} n\to \infty \ ,
\end{align}
asymptotically, and we have
\begin{align}\label{eq:asympratio}
    \frac{\mathcal{F}_g\ui{2n+1}(\xi,t)}{\xi^2 \mathcal{F}_g\ui{2n-1}(\xi,t)} \sim \xi^{-2}\left(\frac{2n-1}{2n+1}\right)^{\beta+1}  \text{~~as~~} n\to \infty\ ,
\end{align}
which approach $\xi^{-2}<1$ as $n\to \infty$ and indicates the divergence of the series with small/medium $\xi$.

Apparently, this oversimplified argument does not prove the asymptotic behaviors of the gluon GPD $F_g(x,\xi,t)$ for large $\xi$ which would require more concrete information on GPDs beyond the scope of this work. In the following subsection, we will explore the large-$\xi$ behavior of GPDs through a GPD model. 

\subsection{Numerical examination of asymptotic expansion through a GPD model}

The above heuristic arguments can be examined by numerical calculations with certain GPD models. Here we consider a simple GPD parameterization model, the double distribution method, which writes the GPDs as~\cite{Radyushkin:1998bz,Radyushkin:1998es}
\begin{align}
H_g(x,\xi,t)&=&H_g^{DD}(x,\xi,t)+|\xi|\theta(|\xi|-|x|)D_g(x,\xi,t)\ ,\\
E_g(x,\xi,t)&=&E_g^{DD}(x,\xi,t)-|\xi|\theta(|\xi|-|x|)D_g(x,\xi,t)\ ,
\end{align}
where $\left\{H,E\right\}^{DD}$ represent the double distribution (DD) terms and $D_g(x,\xi,t)$ is commonly called the D-term.

The DD terms can be written in terms of the integral transformation of double distributions as~\cite{Radyushkin:1998bz,Radyushkin:1998es,Koempel:2011rc}
\begin{eqnarray}
H_{g}^{DD}(x,\xi,t)&=&\int_{-1}^1 \text{d}\beta \int_{-1+|\beta|}^{1-|\beta|}\text{d}\alpha\delta(x-\beta-\alpha\xi)\nonumber\\
&&\times  \pi_g(\alpha,|\beta|) 
|\beta|f_g(|\beta|,t)\ , \\
E_g^{DD}(x,\xi)&=&\int_{-1}^1\text{d}\beta \int_{-1+|\beta|}^{1-|\beta|}\text{d}\alpha\delta(x-\beta-\alpha\xi)
\nonumber\\
&&\times  \pi_g(\alpha,|\beta|) 
e_g(|\beta|,t)\ ,
\end{eqnarray}
where the $\pi_g(\alpha,\beta)$ is the so-called profile function for a double distribution parameterization, and for the gluon GPD it is commonly chosen to be
\begin{equation}
 \pi_g(\alpha,|\beta|)=\frac{15}{16}\frac{((1-|\beta|)^2-\alpha^2)^2}{(1-|\beta|)^5}\ .
\end{equation}
The $f_g(x,t)$ and $e_g(x,t)$ are the $t$-dependent PDFs which correspond to the gluon GPDs $H_g(x,\xi,t)$ and $E_g(x,\xi,t)$ in the semi-forward limit $\xi\to0$. Accordingly, they are subject to the constraints from the forward gluon PDF $g(x)$, e.g., $f_g(x,t=0) = g(x)$. In the following, we will mostly focus on the $H_g$ GPD without loss of generality, where similar arguments should apply to the $E_g$ GPD of which the endpoint constraint exists as well. Taking the gluon PDF from the CT18 global analysis~\cite{Hou:2019efy}, we have
\begin{equation}
g(x)=a_0 x^{a_1-1} (1-x)^{a_2} P_a^g(y)\ ,
\end{equation}
with $y\equiv \sqrt{x}$ and
\begin{align}
\begin{split}
    P_a^g(y)=&\sinh{a_3} (1-y)^3 + \sinh{a_4} 3 y(1-y)^2 \\
    &+a_5 3y^2(1-y)+y^3 ,
\end{split}
\end{align}
where the parameters $a_i$ are presented in Ref.~\cite{Hou:2019efy}. These parameters  together with the profile function $\pi_g(\alpha,\beta)$ provide a model of the $x$- and $\xi$-dependence of the DD term of the gluon GPD $H_{g}^{DD}(x,\xi,t=0)$, whereas the $t$-dependence is not of concern here for the large-$\xi$ analysis.

The D-terms acquire their name for a different reason that they generate the highest power of $\xi$ in the Mellin moments of GPDs, which are the $C_g(t)$ or $D_g(t)$ gravitational form factors depending on the conventions in the case of leading moments~\cite{Ji:1996ek,Ji:1998pc}.\footnote{We do not define separate notations for the GFF $D_g(t)$ and the D-term $D_g(x,\xi,t)$ which are distinguished by their different arguments.} These terms have support solely in the DA-like region $|x|<\xi$, and thus can be parameterized in terms of a set of polynomials of $z\equiv x/\xi$ that are complete on $[-1,1]$. The Gegenbauer polynomials are commonly chosen as they are multiplicatively renormalizable to the leading order of QCD evolution~\cite{Belitsky:1997pc}, with which the D-terms can be written as,
\begin{equation}
\label{eq:Dterm}
    D_g(x,\xi,t)=\frac{3}{2}(1-z^2)^2\sum_{n=1,\rm{odd}}d_g\ui n(t)C_{n-1}\ui{{5}/{2}}(z) \ .
\end{equation}
The $d_g\ui n(t)$s correspond to combinations of the GFF $D_g(t)$ and other generalized form factors in the highest power of $\xi$, and they parameterize the D-term.

Even though the D-terms are generally non-zero and crucial for the polynomiality condition of the GPDs~\cite{Ji:1996nm,Ji:1998pc}, we will not consider them here for the analysis of asymptotic expansion. Since their contributions to the $G(\xi,t)$ can be written explicitly with eq. (\ref{eq:Dterm}) as
\begin{align}
    \frac{1}{2\xi}\int_{-\xi}^\xi \text{d}x \mathcal{A}(x,\xi) \xi  D_g(x,\xi,t)=2\sum_{n=1,\rm{odd}} d_g\ui n(t)\ ,
\end{align}
they generate zero imaginary part and the real part converges as long as $d\ui{n+2}_g(t)<d_g\ui n(t)$ as $n\to \infty$, which is generally assumed to be true unless the higher moments anomalously increase. Therefore, we will focus on the asymptotic behaviors of the DD-terms.

Given the above GPD model, we now consider the hadronic matrix element $G(\xi,t)$. Summing/averaging over all the final/initial proton polarizations in the unpolarized case, one has the squared hadronic matrix element as
\begin{align}\label{eq:Gwigcffs}
\begin{split}
    \left|G(\xi,t)\right|^2=&\Big[\left(1-\xi^2\right)\left|\mathcal{H}_{g\rm{C}}\right|^2-2\xi^2\text{Re}\left[\mathcal{H}_{g\rm{C}}^* \mathcal{E}_{g\rm{C}}\right] \\
        & -\left(\xi^2+\frac{t}{4M_p^2}\right)\left|\mathcal{E}_{g\rm{C}}\right|^2\Big]\ ,
\end{split}
\end{align}
where the gluonic Compton form factor (gCFF) $\mathcal{H}_{g\rm{C}}$  is defined as,
\begin{align}\label{eq:gcffs}
\mathcal{H}_{g\rm{C}}(\xi,t)\equiv\frac{1}{2\xi} \int_{-1}^1\text{d}x{\cal A}(x,\xi)H_g(x,\xi,t) \ ,
\end{align}
and similarly the $\mathcal{E}_{g\rm{C}}$. Here we consider the GPD $H_g$ and gCFF $\mathcal{H}_{g\rm{C}}(\xi,t)$ as an example without loss of generality. 

Both $\mathcal{H}_{g\rm{C}}(\xi,t)$ and $\mathcal{E}_{g\rm{C}}(\xi,t)$ are complex although the GPDs are real, since the Wilson coefficient $\mathcal{A}(x,\xi)$ is complex. The real part of the gCFF $\mathcal{H}_{g\rm{C}}(\xi,t)$ can be written as a principal-value integral of the Wilson coefficient $\mathcal{A}(x,\xi)$ and the gluon GPD $H_g(x,\xi,t)$ that reads,
\begin{align}\label{eq:gcffre}
\begin{split}
    \text{Re}\mathcal{H}_{g\rm{C}}(\xi,t)&=\int_{-1}^1\text{d}x~\text{P.V.}~\frac{1}{\xi^2-x^2}H_g(x,\xi,t) \\
    &= \sum_{n=0}^{\infty}\frac{2}{\xi^{2n+2}} \mathcal{H}_g\ui{2n+1}(\xi,t)\ ,
\end{split}
\end{align}
where P.V. stands for taking the principal value and $\mathcal{H}_g\ui{2n+1}(\xi,t)$ are the Mellin moments of $H_g(x,\xi,t)$ GPD defined similarly to eq.(\ref{eq:MelMom}), whereas the imaginary part of the gCFF can be written with the GPD at the crossover line $x=\pm \xi$ as
\begin{align}\label{eq:gcffim}
\begin{split}
    \text{Im}\mathcal{H}_{g\rm{C}}(\xi,t)=&\frac{\pi}{2\xi} \left[H_g(\xi,\xi,t)+H_g(-\xi,\xi,t) \right]\\
    &=\frac{\pi}{\xi} H_g(\xi,\xi,t)\ .
\end{split}
\end{align}
As just mentioned, since the D-terms vanish at $x=\pm \xi$ and generate zero imaginary part, the imaginary part of the gCFF will be from the DD terms only.

\begin{figure}[t]
    \includegraphics[width=0.48\textwidth]{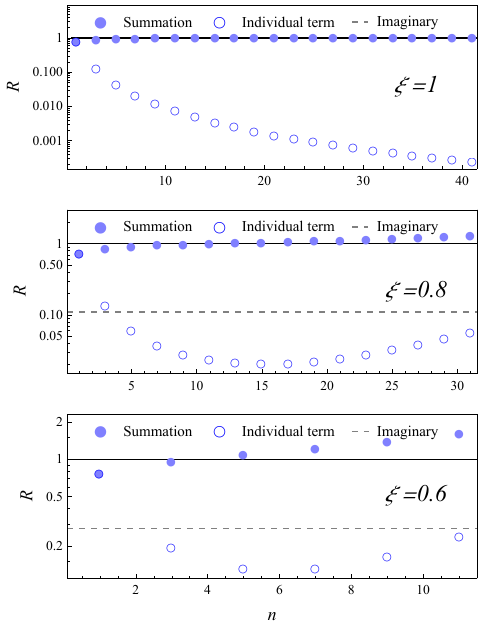}
    \caption
    {\raggedright The asymptotic behavior of the $R$ ratios for different values of $\xi$. The solid line stands for the $R=1$ line which the real part of $R$ should converge to, the dashed line stands for the $r_{\rm{Im}}(\xi)$. The open blue dots are the contribution from moments of different $n$ whereas the solid blue dots are the partial sums of them.}
    \label{fig:asymp}
\end{figure}

In FIG. \ref{fig:asymp} we show the asymptotic behavior of such expansions for different values of $\xi$ where we define
\begin{equation}
    R\equiv \mathcal{H}_{g\rm{C}}(\xi)/\text{Re}\mathcal{H}_{g\rm{C}}(\xi)=1 +i r_{\rm{Im}}(\xi)\ .
\end{equation}
For discussion, we set $t=0$ and all suppressed $t$ should be considered to be zero in this section, although we note that there could be non-trivial entanglements of $x,\xi$ and $t$ in general.\footnote{For instance, one might expect the higher moments of GPDs to be typically harder, i.e., they are associated with larger pole masses and thus flatter $t$-slopes. Therefore, the $j$th moment is enhanced by a factor of $(M_j^2/M_1^2)^{p}$ compared to the 1st moment with $p=2$ or 3 for dipole or tripole etc. at large $t$. This enhancement competes with the higher-moment suppression as $t$ gets large, which, however, cannot be quantitatively examined at this point. } The solid line stands for the $R=1$ line which the real part of $R$ should approach if the series converges, and the dashed line stands for the $r_{\rm{Im}}(\xi)$ which approaches zero in the $\xi \to 1$ limit according to the endpoint constraint:
\begin{equation}\label{eq:endpoint}
    \text{Im}\mathcal{H}_{g\rm{C}}(\xi,t)=\frac{\pi}{\xi}H_g(\xi,\xi,t) \to 0 \text{~~as~~} \xi \to 1\ .
\end{equation}
The open blue dots are the individual contributions from moments of different $n$, whereas the solid dots are the partial sums of them.

\begin{figure}[t]
    \includegraphics[width=0.48\textwidth]{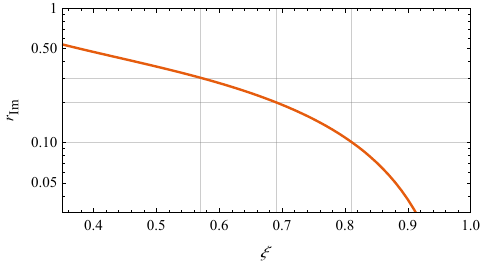}
    \caption
    {\raggedright The imaginary ratio $r_{\rm{Im}}(\xi)$ as a function of $\xi$. It approaches zero when $\xi \to 1$, resulting from the endpoint constraints.}
    \label{fig:rimplt}
\end{figure}

The real part of the gCFF $\mathcal{H}_{g\rm{C}}(\xi,t)$ does behave as expected from the plot. At $\xi=1$, the series appears convergent even when summing over the very high moments, a behavior likely model-independent. By contrast, for $\xi<1$ although the partial sums start to look convergent for the lower moments, they ultimately diverge when higher moments are involved. In addition, a truncation at the minimal term always leads to $R\approx 1$ which is known as the superasymptotic approximation. As $\xi$ decreases, the asymptotic expansion will diverge earlier and faster, and eventually cease to work. On the other hand, the behavior of the imaginary part is more obscure apart from their vanishing behavior at $\xi\to1$. As shown in FIG. \ref{fig:asymp}, the imaginary ratio $r_{\rm{Im}}(\xi)$ (dashed line) indeed vanishes in the top $\xi=1$ plot, and then increases as $\xi$ gets lower. The full $\xi$-dependence of $r_{\rm{Im}}(\xi)$ is also shown in FIG. \ref{fig:rimplt}. A more careful treatment to take these imaginary parts into account could be to parameterize and include them in analyses as well. However, we will focus on the real parts in this work.

Therefore, we comment that the gCFFs $\mathcal{H}_{g\rm{C}}(\xi,t)$ and $\mathcal{E}_{g\rm{C}}(\xi,t)$ can be approximated by their leading Mellin moments in the sense of an asymptotic expansion in the $\xi \to 1^-$ limit. For relatively lower $\xi$, there are two main corrections from the higher moments' contributions and the non-vanishing imaginary part.

\subsection{Asymptotic form of the gCFFs}

The asymptotic expansion of the gCFFs $\mathcal{H}_{g\rm{C}}(\xi,t)$ and $\mathcal{E}_{g\rm{C}}(\xi,t)$ provides a rather effective tool to study the behavior of them in the $\xi\to 1^-$ limit. Besides the leading-moment approximation, we could also obtain other useful information in a larger region of $\xi<1$. To show this, consider the polynomiality conditions of the gluon GPDs that read~\cite{Ji:1998pc}
\begin{align}
\begin{split}
    \label{eq:GPDmoments}
    \mathcal{H}\ui{2n+1}_g(\xi,t)= &\sum_{i=0}^n(2\xi)^{2i}A_g\ui{2n+2,2i}(t) \\
    &\quad+(2\xi)^{2n+2}C_g\ui{2n+2}(t) \ ,
\end{split} \\
\begin{split}
    \mathcal{E}\ui{2n+1}_g(\xi,t)= &\sum_{i=0}^n(2\xi)^{2i}B_g\ui{2n+2,2i}(t) \\
    &\quad-(2\xi)^{2n+2}C_g\ui{2n+2}(t) \ ,
\end{split}
\end{align}
which require that the Mellin moments of the GPDs, the $\mathcal{H}\ui{2n+1}_g(\xi,t)$ and $\mathcal{E}\ui{2n+1}_g(\xi,t)$ here, must be finite-order polynomials of $\xi$. The coefficients in these polynomials, the $A_g\ui{2n+2,2i}(t)$, $B_g\ui{2n+2,2i}(t)$, and $C_g\ui{2n+2}(t)$, are known as the generalized form factors. We note that when $n=0$ these generalized form factors correspond to the well-known gluonic GFFs~\cite{Ji:1996ek}: $A_g(t)=  A\ui{2,0}_g(t)$, $B_g(t)=  B\ui{2,0}_g(t)$, and $C_g(t)=  C\ui{2}_g(t)$, which will be used to avoid redundant indices. 

Plugging them back into eq. (\ref{eq:gcffre}), we obtain the asymptotic behaviors of the real parts of gCFFs as presented in the introduction:
\begin{align}
\begin{split}
    \text{Re}\mathcal{H}_{g\rm{C}}(\xi,t)&= \mathdutchcal{C}_g(t) + \xi^{-2} \mathdutchcal{A}_g\ui2(t)+\xi^{-4} \mathdutchcal{A}_g\ui4(t)+\cdots\ ,
\end{split}\\
\begin{split}
    \text{Re}\mathcal{E}_{g\rm{C}}(\xi,t)&= -\mathdutchcal{C}_g(t) + \xi^{-2} \mathdutchcal{B}_g\ui2(t)+\xi^{-4} \mathdutchcal{B}_g\ui4(t)+\cdots\ ,
\end{split}
\end{align}
where each of the new coefficients $\mathdutchcal{A}_g\ui{2n}(t)$, $\mathdutchcal{B}_g\ui{2n}(t)$, and $\mathdutchcal{C}_g(t)$ are given by those generalized form factors as
\begin{align}\label{eq:acoesum}
\begin{split}
    \mathdutchcal{A}_g\ui{2n}(t) \equiv \sum_{k=0}^\infty 2^{2k+1} A\ui{2k+2n,2k}_g(t)\ ,
\end{split}\\
\begin{split}\label{eq:bcoesum}
    \mathdutchcal{B}_g\ui{2n}(t) \equiv \sum_{k=0}^\infty 2^{2k+1} B\ui{2k+2n,2k}_g(t)\ ,
\end{split}\\
\begin{split}\label{eq:ccoesum}
    \mathdutchcal{C}_g(t) \equiv \sum_{k=0}^\infty 2^{2k+3} C\ui{2k+2}_g(t)\ ,
\end{split}
\end{align}
where $n$ is a positive integer. Each of the series expansions here should similarly be understood as an asymptotic expansion that contains infinite terms. The coefficients $\mathdutchcal{A}_g\ui{2n}(t)$ and $\mathdutchcal{B}_g\ui{2n}(t)$ contain moments of order $2n+2k$ for non-negative integers $k$ according to eqs. (\ref{eq:acoesum}) and (\ref{eq:bcoesum}). Thus, higher-order coefficients with $n\ge2$ contain higher order moments with $2n\ge 4$, and $\mathdutchcal{A}_g\ui{2}(t)$, $\mathdutchcal{B}_g\ui{2}(t)$ and $\mathdutchcal{C}_g(t)$ are the three coefficients that contain leading moments, namely the gluonic GFFs. Therefore, the behavior of the real part of the gCFFs could be approximated by these new coefficients as
\begin{align}
\begin{split}\label{eq:hreasymp}
    \text{Re}\mathcal{H}_{g\rm{C}}(\xi,t)&\approx \mathdutchcal{C}_g(t) + \xi^{-2} \mathdutchcal{A}_g\ui2(t)\ ,
\end{split}\\
\begin{split}\label{eq:ereasymp}
    \text{Re}\mathcal{E}_{g\rm{C}}(\xi,t)&\approx -\mathdutchcal{C}_g(t) + \xi^{-2} \mathdutchcal{B}_g\ui2(t)\ ,
\end{split}
\end{align}
where higher-order coefficients are dropped. In the leading-moment approximation, one has 
\begin{eqnarray}\label{eq:leadingMelMomA}
    \mathdutchcal{A}_g\ui{2}(t)&\approx 2 A_g(t)\ ,\\\label{eq:leadingMelMomB}
    \mathdutchcal{B}_g\ui{2}(t)&\approx 2 B_g(t)\ ,\\\label{eq:leadingMelMomC}
    \mathdutchcal{C}_g(t)&\approx 8 C_g(t)\ ,
\end{eqnarray}
 whereas the other coefficients are just 0. Then the formula reduces to the previous one in Ref. \cite{Guo:2021ibg}. 

\begin{figure}[t]
    \includegraphics[width=0.48\textwidth]{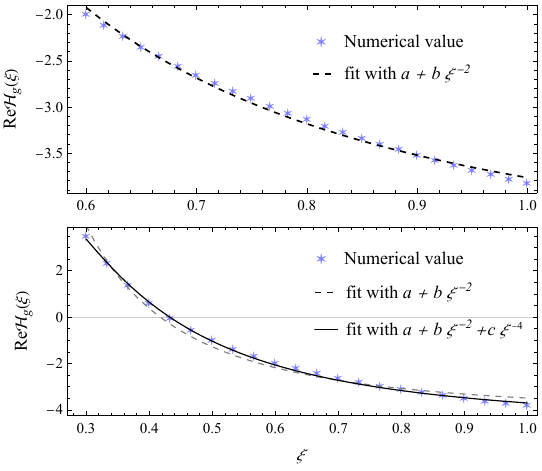}
    \caption
    {\raggedright The $\xi$-dependence of the $\text{Re}\mathcal{H}_{g\rm{C}}(\xi,0)$ calculated with the above DD model. When only large $\xi>0.6$ (up) is considered, the $\text{Re}\mathcal{H}_{g\rm{C}}$ can be fitted well with just two terms, whereas an extra $\xi^{-4}$ term is needed when including the medium $\xi>0.3$ as well (down).}
    \label{fig:xiscale}
\end{figure}

In FIG. \ref{fig:xiscale}, we show the $\xi$-dependence of the $\text{Re}\mathcal{H}_{g\rm{C}}(\xi,t)$ at $t=0$ calculated with the same DD GPD model as above except that a D-term with $d_g^{(1)}(0)=5/4 D_g(0)$ is added so that the $\mathdutchcal{C}_g(t)$ coefficient will be non-zero. We take $D_g(0)=-1.93$ from the lattice simulation of gluonic GFFs~\cite{Pefkou:2021fni} as the input.  In the upper plot, the $\xi$-dependence of the numerical $\text{Re}\mathcal{H}_{g\rm{C}}(\xi,t)$ is shown to be approximated well by the form $a+b~\xi^{-2}$ when $\xi>0.6$, consistent with the asymptotic form above. As $\xi$ gets smaller, e.g., for $\xi>0.3$ as shown in the lower plot, the higher-order coefficients get more relevant and an extra $\xi^{-4}$ term is needed. It looks that the asymptotic form holds even at relatively low $\xi$ where the leading-moment approximation may not. However, this could depend on the GPD model.

It is interesting to note that numeric calculations lead to the following behavior for ${\rm Re}{\cal H}_{g\rm{C}}(\xi,t=0)$ at large $\xi$: $\lim_{\xi\to 1}{\rm Re}{\cal H}_{g\rm{C}}(\xi,0)\approx 5/4\left(2A_g(0)+2D_g(0)\right)$. This may come from the particular parameterization of the double distribution part of the GPD gluon distributions used in our analysis. This feature will be discussed in more details in the next subsection.

Thus, we obtain the asymptotic forms of the real parts of the gCFFs with a set of coefficients to be determined. Although the gCFFs also have imaginary parts, they vanish in the large $\xi$ limit. Therefore, in the large or even medium-$\xi$ region, such forms are supposed to describe the experimental measurements well and allow us to extract these coefficients from the data. This will be discussed with more details in the next section.

\subsection{Reconstruction of leading moments and conformal moment expansion}

Before moving on to the data, we discuss the last step in the extraction of the gluonic GFFs --- suppose these coefficients
$\mathdutchcal{A}_g\ui{2}(t)$, $\mathdutchcal{B}_g\ui{2}(t),\mathdutchcal{C}_g(t) \cdots$ are reliably extracted with the
asymptotic forms of the gCFFs $\mathcal{H}_{g\rm{C}}(\xi,t)$ and $\mathcal{E}_{g\rm{C}}(\xi,t)$, can we reconstruct the leading moments, the gluonic GFFs, from these coefficients? In the leading-moment approximation, one could simply use eqs. (\ref{eq:leadingMelMomA})--(\ref{eq:leadingMelMomC}) and take $A_g(t)\approx 1/2\mathdutchcal{A}_g\ui{2}(t)$, $B_g(t) \approx 1/2  \mathdutchcal{B}_g\ui{2}(t) $, and $  C_g(t) \approx 1/8 \mathdutchcal{C}_g(t)$ noting that there are corrections from higher moments. In this subsection, we will consider such reconstructions more carefully.

We start with the example in the previous subsection shown in FIG. \ref{fig:xiscale}. Suppose these numerical $\text{Re}\mathcal{H}_{g\rm{C}}(\xi,0)$ were obtained with experimental measurements with infinite precision in the ideal limit, we will consider if and how the leading moments can be reconstructed. With the asymptotic form in eq. (\ref{eq:hreasymp}) of the $\text{Re}\mathcal{H}_{g\rm{C}}(\xi,0)$, we consider the simplest two-term fit in the form of $a + b~\xi^{-2}$ for $\xi>0.6$ in the upper plot, which gives
\begin{equation}
    \text{Re}\mathcal{H}_{g\rm{C}}(\xi,0) =-4.81 +1.04 ~\xi^{-2} \qquad (\xi>0.6)\ .
\end{equation}
On the other hand, for a larger range of $\xi>0.3$ in the lower plot, we consider instead a three-term fit in the form $a + b~\xi^{-2} + c ~\xi^{-4}$ and obtain
\begin{equation}\label{eq:xi3termfit}
    \text{Re}\mathcal{H}_{g\rm{C}}(\xi,0) =-4.70 +1.02 ~\xi^{-2}-0.03~\xi^{-4} \quad (\xi>0.3)\ .
\end{equation}
Since the $\xi>0.3$ and $\xi>0.6$ data are generated with the same GPD, the similar extracted coefficients for the constant and $\xi^{-2}$ terms in the two fits reflect the reliability of the extraction. On the other hand, the $\xi^{-4}$ term in the three-term fit has a small coefficient, suggesting that the asymptotic form still holds and the terms from higher moments are still suppressed, although as noted above, this could depend on the GPD model.

We also consider a two-term fit to the $\xi>0.3$ data, which will not fit the data as well and the extracted coefficients deviate from the large-$\xi$ ones:
\begin{equation}\label{eq:xi2termfit}
    \text{Re}\mathcal{H}_{g\rm{C}}(\xi,0) =-4.23 +0.74 ~\xi^{-2} \qquad (\xi>0.3)\ .
\end{equation}
Such results suggest that even when the asymptotic form holds, the leading-moment approximation could still receive sizable corrections, particularly for lower $\xi$.
Therefore, it will be more reliable to first use the asymptotic form to extract the leading-order coefficients $\mathdutchcal{A}_g\ui{2}(t)$, $\mathdutchcal{B}_g\ui{2}(t),$ and $\mathdutchcal{C}_g(t)$ from the measured gCFFs rather than to apply the leading-moment approximation directly to the extraction.

With that in mind, we take the extracted values of the two coefficients $\mathdutchcal{A}_g\ui{2}(0)$ and $\mathdutchcal{C}_g(0)$ to be around $1.04$ and $-4.8$ respectively, based on the numerical gCFFs $\mathcal{H}_{g\rm{C}}(\xi,t)$. Once these coefficients are extracted, the GFFs $A_g(0)$ and $C_g(0)$ can be reconstructed in the leading-Mellin-moment approximation to be
\begin{equation}
\label{eq:coefextract}
    \frac{1}{2}\mathdutchcal{A}_g\ui2(0)\approx 0.52 \text{~~and~~} \frac{1}{8}\mathdutchcal{C}_g(0)\approx -0.60\ ,
\end{equation}
which are, however, larger than the input values of them:
\begin{equation}\label{eq:acinput}
    A_g(0)\approx0.385 \text{~~and~~} C_g(0)\approx -0.48\ .
\end{equation}
The differences are caused by the higher moments in $\mathdutchcal{A}_g\ui{2}(0)$ and $\mathdutchcal{C}_g(0)$ according to eqs. (\ref{eq:acoesum}) and (\ref{eq:ccoesum}). To make things worse, such differences persist in the $\xi\to 1$ limit: the leading moments $A_g(0)$ or $C_g(0)$ take up only about $80\%$ of the total contributions, causing a consistent systematical uncertainty of about $25\%$ in the extraction under leading-moment approximation even at $\xi=1$.

While this discrepancy between the extracted and input leading moments should be mitigated when explicitly including the higher moments, we note that these higher-order Mellin moments are typically even harder to obtain, making the attempt to separate their contributions from the leading ones rather unrealistic. Instead, we consider a method that does not require their explicit values to improve the extraction --- suppose we can rearrange the asymptotic series such that the leading terms get more relevant, then the leading moment approximation will work better. Accordingly, we consider another expansion of the GPDs besides the Mellin moments expansion, which is known as the conformal moment expansion.

We will not fully go through the conformal moment expansion of GPDs, which has been systematically studied in Ref.~\cite{Mueller:2005ed}. Simply speaking, the conformal moment expansion projects the GPD onto a complete set of Gegenbauer polynomials $C_{n}\ui{\lambda}(x/\xi)$ and expresses the GPD as the formal sum of these Gegenbauer polynomials with $\lambda =5/2$ for the gluon. So, we have 
\begin{equation}
\label{eq:confexp}
    F_g(x,\xi,t)=-\sum_{\substack{j=1\\ \rm{odd}}}^\infty \left(p_{g,j}(x,\xi)+p_{g,j}(-x,\xi) \right)\mathcal{F}_{g,j}^{\rm{conf}}(\xi,t) \ ,
\end{equation}
and the $j$-th conformal moment $\mathcal{F}_{g,j}^{\rm{conf}}(\xi,t)$ is defined as 
\begin{equation}
\label{eq:confdef}
   \mathcal{F}_{g,j}^{\rm{conf}}(\xi,t)\equiv \int_{-1}^1\text{d} x ~c_{g,j}(x,\xi)F_g(x,\xi,t)\ ,
\end{equation}
where $p_{g,j}(x,\xi)$ and $c_{g,j}(x,\xi)$ are known functions expressible in terms of the Gegenbauer polynomials $C_{j-1}^{(5/2)}(x/\xi)$ which can be found in Ref.~\cite{Mueller:2005ed}.

With the conformal moment expansion, one can show that the gCFFs can be written in terms of the corresponding conformal moments as
\begin{align}\label{eq:gcffconf}
\begin{split}
    \text{Re}\mathcal{H}_{g\rm{C}}(\xi,t) =2\sum_{\substack{j=1\\ \rm{odd}}}^\infty \xi^{-j-1} \mathcal{A}_j^{\rm{conf}} \mathcal{F}_{g,j}^{\rm{conf}}(\xi,t)\ ,
\end{split}
\end{align}
where $\mathcal{A}_j^{\rm{conf}}$ is the Wilson coefficient in the conformal moment space that reads~\cite{Mueller:2005ed}:
\begin{equation}
    \mathcal{A}_j^{\rm{conf}} =\frac{2^{j+2}\Gamma\left(5/2+j\right)}{\Gamma\left(3/2\right)\Gamma\left(4+j\right)}\ .
\end{equation}
\begin{figure}[t]
    \includegraphics[width=0.48\textwidth]{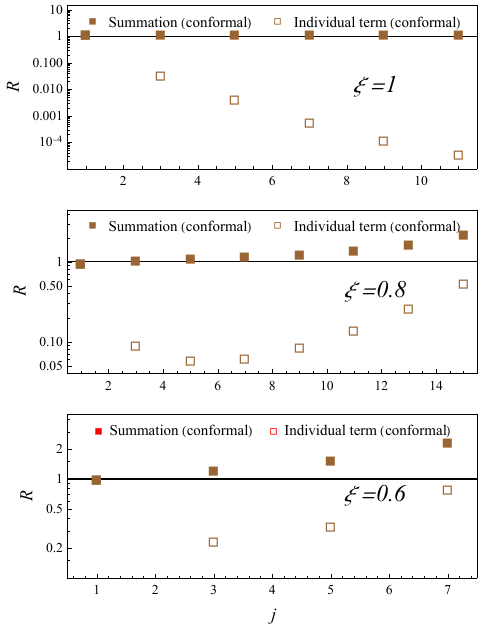}
    \caption
    {\raggedright The asymptotic behavior of the $R$ ratios for different values of $\xi$ based on the conformal moment expansion. The solid line stands for the $R=1$ line which the real part of $R$ should converge to. The open brown dots are the contribution from moments of different $j$ whereas the solid brown dots are the partial sums of them.}
    \label{fig:asympconf}
\end{figure}
Then we can equivalently use eq. (\ref{eq:gcffconf}) as an asymptotic expansion of the real part of the gCFFs. 

In FIG. \ref{fig:asympconf} we study the asymptotic behavior of the conformal moment expansions for different values of $\xi$ analogous to what we did with the Mellin moment expansion before. Similar to the previous case, the series seems convergent at $\xi=1$ whereas for $\xi<1$ it behaves as an asymptotic series that reaches the best approximation at the minimal term. On the other hand, the conformal moment expansion appears to converge (and diverge after passing the minimal term) faster compared to the Mellin moment expansion shown in FIG. \ref{fig:asymp}. As we discussed, this can be understood as a reshuffle of the higher moments into the lower ones.

Due to this reshuffle effect, it looks that the contribution from the leading conformal moment dominates over all the other moments in the $\xi \to 1$ limit, i.e., we have
\begin{equation}
    R_1^{\rm{conf}}\equiv \frac{2 \xi^{-2} \mathcal{A}_1^{\rm{conf}} \mathcal{F}_{g,1}^{\rm{conf}}(\xi,t)}{\text{Re}\mathcal{H}_{g\rm{C}}(\xi,t)}  \approx 1\ .
\end{equation}
This also magically holds even for lower $\xi$ as shown in FIG. \ref{fig:asympconf}. In the three plots with $\xi=1,0.8,\text{ and }0.6$,  the relative contributions of the leading conformal moments are $R_1^{\rm{conf}}=1,1.06,\text{ and }1.08$, compared to those of the leading moment moments $R_1^{\rm{Mel}}=0.8,0.85,\text{ and }0.86$. We also note that their ratio is always $5/4$ since 
\begin{equation}
  \mathcal{A}_1^{\rm{conf}}  = \frac{5}{4} \text{~~and~~}   \mathcal{F}_{g,1}^{\rm{conf}}(\xi,t)=\mathcal{F}_{g}^{(1)}(\xi,t)\ ,
\end{equation}
where the relation between the conformal and Mellin moment only holds at $j=1$ though. 

Thus, it appears that the leading-conformal-moment approximation with an extra conformal ratio $\mathcal{A}_1^{\rm{conf}}$ multiplied can be considered as an improvement to the leading-Mellin-moment approximation. Then we have:
\begin{eqnarray}\label{eq:leadingConfMomA}
    \mathdutchcal{A}_g\ui{2}(t)&\approx 2 \mathcal{A}_1^{\rm{conf}} A_g(t)\ ,\\\label{eq:leadingConfMomB}
    \mathdutchcal{B}_g\ui{2}(t)&\approx 2 \mathcal{A}_1^{\rm{conf}} B_g(t)\ ,\\\label{eq:leadingConfMomC}
    \mathdutchcal{C}_g(t)&\approx 8 \mathcal{A}_1^{\rm{conf}} C_g(t)\ .
\end{eqnarray}
Applying this back to the coefficients $\mathdutchcal{A}_g\ui{2}(t)\approx 1.04$ and $\mathdutchcal{C}_g(t)\approx -4.8$ obtained before, the conformal reconstruction of leading moments reads
\begin{equation}
    A_g(0)\approx0.42 \text{~~and~~} C_g(0)\approx -0.48\ ,
\end{equation}
that agrees well with the input values in eq. (\ref{eq:acinput}). Note that the extracted $A_g(0)$ is affected by the higher moments of PDFs, whereas the extracted $C_g(0)$ is not since only one D-term $d_g^{(1)}(0)=5/4 D_g(0)$ was put in. This explains the almost perfectly extracted $C_g(0)$, while the extracted $A_g(0)$ is a bit off but reasonably close.

\subsection{Dispersion relation and applicability to general GPDs}

At the end of the section, we discuss the applicability of the above arguments to more general GPDs. First, we note that the arguments in this section rely mostly on the endpoint constraints of the GPD as well as the analyticity of the Compton-like amplitudes which are generally assumed to be true for all GPDs/amplitudes. 

The analyticity property of the GPDs and gCFFs can be exploited with the dispersion relation~\cite{Drechsel:2002ar,Kumericki:2007sa}, which analytically continues the GPDs and gCFFs to the $\xi>1$ region at fixed $t$. Utilizing Cauchy's integral formula, the gCFF can be expressed in terms of a contour integral along the branch cut above the threshold, and one eventually obtains~\cite{Drechsel:2002ar,Kumericki:2007sa}
\begin{equation}
\label{eq:disp}
    \mathcal{H}_{g\rm{C}}(\xi,t) =\frac{1}{\pi } \int^{\xi_{\rm{th}}}_0 \text{d}\xi' \frac{2\xi' \text{Im}\mathcal{H}_{g\rm{C}}(\xi,t)}{\left(\xi-\xi'-i0\right)\left(\xi+\xi'+i0\right)}   + \mathdutchcal {C}_g(t)\ ,
\end{equation}
where the $\mathdutchcal {C}_g(t)$ is the so-called subtraction term which coincides with the $\mathdutchcal {C}_g(t)$ coefficient defined previously. A similar relation applies to the $\mathcal{E}_{g\rm{C}}(\xi,t)$ with an extra minus sign in the subtraction term. The dispersion relation naturally applies to the region below the threshold, i.e., $\xi>\xi_{\rm{th}}\approx 1>\xi'$. Therefore, it can be expanded as
\begin{equation}
\label{eq:dispmom}
    \text{Re}\mathcal{H}_{g\rm{C}}(\xi,t) =2  \sum_{n=0}^\infty   \xi^{-2n-2} \text{Im}\mathcal{H}_{g\rm{C}}\ui{2n+1}(t)  +\mathdutchcal {C}_g(t)\ ,
\end{equation}
where $\text{Im}\mathcal{H}_{g\rm{C}}\ui{n}(t)$ is defined as the $n$th moment of the imaginary part of the amplitude by
\begin{align}
\begin{split}
    \label{eq:gCFFMel}
    \text{Im}\mathcal{H}_{g\rm{C}}^{n}(t) &\equiv \frac{1}{\pi}\int_0^1 \text{d} \xi \xi^{n}\text{Im}\mathcal{H}_{g\rm{C}}(\xi,t)\\&= \int_0^1 \text{d} \xi \xi^{n-1} H_g(\xi,\xi,t)\ .
\end{split}
\end{align}
Comparing it with the previous results, one obtains the matching condition
\begin{equation}
     \text{Im}\mathcal{H}_{g\rm{C}}\ui{2n+1}(t)  =  \sum_{k=0}^{\infty} 2^{2k} A_g\ui{2n+2k+2,2k}(t) =\frac{1}{2} \mathdutchcal {A}_g\ui{2n+2}(t) \ ,
\end{equation}
where similar relations in terms of the conformal moments are presented in Ref.~\cite{Kumericki:2007sa}. Thus, the asymptotic expansion is consistent with the dispersion relation and our coefficients $\mathdutchcal {A}_g\ui{2n}(t)$ and $\mathdutchcal {B}_g\ui{2n}(t)$ can also be regarded as the Mellin moments of the imaginary part of the gCFFs. Note that the Mellin moments of the imaginary gCFF should be distinguished from the those of the GPDs themselves --- the imaginary gCFFs correspond to GPDs at $x=\xi$, while for GPDs $x$ and $\xi$ are uncorrelated.

Then, we could have very similar arguments for the asymptotic behaviors of the real part of the gCFF utilizing the endpoint constraint of the imaginary part of the gCFFs. Assuming that the imaginary part of the gCFFs approaches zero according to $(1-\xi)^\alpha$ when $\xi \to 1$ due to endpoint constraint that it vanishes at $\xi=1$, its Mellin moments at large $n$ will be asymptotically
\begin{equation}
     \text{Im}\mathcal{H}_{g\rm{C}}\ui{n}(t)  \sim n^{-\alpha-1} \text{~~as~~} n\to \infty \ ,
\end{equation}
and hence we have 
\begin{align}
    \frac{\text{Im}\mathcal{H}_{g\rm{C}}\ui{2n+1}(t)}{\xi^2 \text{Im}\mathcal{H}_{g\rm{C}}\ui{2n-1}(t)} \sim \xi^{-2}\left(\frac{2n-1}{2n+1}\right)^{\alpha+1}  \text{~~as~~} n\to \infty\ .
\end{align}
This applies to the $\xi \to 1^-$ case, different from the near-forward arguments at the beginning of this section. Consequently, the asymptotic expansion is expected in the $\xi \to 1^-$ limit.

To examine the parameterization dependence of the large-$\xi$ behavior of GPDs, we also go through the above analysis with another parameterization of GPDs based on conformal moments~\cite{Guo:2022upw,Guo:2023ahv}. We observe similar behaviors in the large $\xi\to 1^-$ limit, though in the medium/lower $\xi$ region, the higher-moment contamination appears stronger due to the larger higher moments in this parameterization. Thus, one should be more careful of the applicability to the medium/lower $\xi$ region. Besides, the leading-conformal-moment approximation still works better than the leading-Mellin-moment approximation with this parameterization, although a rigorous proof of such statements seems improbable with only the endpoint constraints.

Furthermore, we note that it is well-known that there is the so-called inverse problem that GPDs cannot be uniquely determined by the Compton-like amplitudes. This inverse problem is reflected in eqs. (\ref{eq:hreasymp}) and (\ref{eq:ereasymp}) that the gCFFs contain in principle infinite coefficients $\mathdutchcal{A}_g\ui{2n}(t)$ where each of them contains infinite generalized form factors. However, the asymptotic expansion allows one to extract the dominant coefficients such as $\mathdutchcal{A}_g\ui{2}(t)$, $\mathdutchcal{B}_g\ui{2}(t)$, and $\mathdutchcal{C}_g(t)$ from the gCFFs in the large-$\xi$ limit, distinguished from the general case where the $\xi$-dependence is not known. Then the leading moments can be effectively extracted with these coefficients, while the higher-moment contamination will enter the systematical uncertainties. Lattice calculations of GPD moments have shown the suppression of higher moments up to the 5th moments for the $A\ui{n,0}_{q}(t)$ and $B\ui{n,0}_{q}(t)$ form factors~\cite{Bhattacharya:2023ays}, while similar behaviors could be expected for the other generalized form factors including the gluonic ones. These results provide extra support for the leading-moment approximation.

\section{Analysis of the differential cross section of the near-threshold \texorpdfstring{$J/\psi$}{JPSI} photo-production}

\label{sec:dataanaly}

\begin{figure}[t]
    \includegraphics[width=0.48\textwidth]{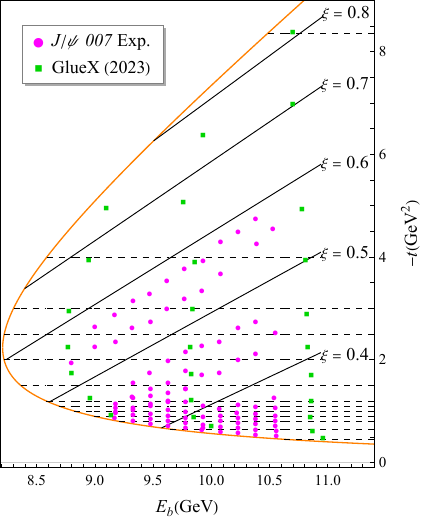}
    \caption
    {\raggedright Differential cross-section data points on the $(E_b,-t)$ plane. Each dot represents a data point from the $J/\psi$ 007 experiment at JLab Hall C (circle)~\cite{Duran:2022xag} or GlueX collaboration at Hall D (square)~\cite{GlueX:2023pev}. Solid lines correspond to contours of equal $\xi$ while the data points with relatively close $|t|$ will be binned together according to the dashed line for comparisons.}
    \label{fig:jpsixitplt}
\end{figure}

Until now, we have studied the large-$\xi$ properties of gCFFs based on the endpoints behaviors of the gluonic GPDs. We find that in the large $\xi\to 1^-$ limit, the real parts of the gCFFs can be written as an asymptotic series in powers of $\xi$, whereas the imaginary parts would vanish and serve as higher order corrections only. Then, these asymptotic behaviors of the gCFFs will also be reflected in the differential cross sections of, for instance, the near-threshold $J/\psi$ photo-production, which will be further investigated in this section.

With the asymptotic forms of the real parts of the gCFFs $\text{Re}\mathcal{H}_{g\rm{C}}$ and $\text{Re}\mathcal{E}_{g\rm{C}}$ as in eqs. (\ref{eq:hreasymp}) and (\ref{eq:ereasymp}), we obtain the asymptotic form of the squared hadronic matrix element $G(\xi,t)$ given in eq. (\ref{eq:Gwigcffs}) as,
\begin{align}\label{eq:xsecscale}
\begin{split}
    \left|G(\xi,t)\right|^2=&\xi^{-4} \Big[G_{0}(t)+\xi^{2} G_{2}(t)+\xi^{4} G_{4}(t)\Big]+\cdots\ ,
\end{split}
\end{align}
where $\cdots$ stands for terms associated with the higher-order coefficients $\mathdutchcal{A}\ui4_g(t)$, $\mathdutchcal{B}\ui4_g(t)$ and the imaginary gCFFs $\text{Im}\mathcal{H}_{g\rm{C}}(\xi,t)$ and $\text{Im}\mathcal{E}_{g\rm{C}}(\xi,t)$ that are not considered for now. The coefficients of different $\xi$-scaling terms can be written in terms of the coefficients $\mathdutchcal{A}_g\ui{2}(t)$, $\mathdutchcal{B}_g\ui{2}(t)$, and $\mathdutchcal{C}_g(t)$ as 
\begin{align}\label{eq:g0exp}
\begin{split}
G_0(t) =& \left(\mathdutchcal{A}\ui2_g(t)\right)^2-\frac{t}{4M_N^2}\left(\mathdutchcal{B}\ui2_g(t)\right)^2\ ,
\end{split}\\\label{eq:g2exp}
\begin{split}
G_2(t) =&2\mathdutchcal{A}\ui2_g(t)\mathdutchcal{C}_g(t)+2\frac{t}{4M_N^2}\mathdutchcal{B}\ui2_g(t)\mathdutchcal{C}_g(t)\\
& - \left(\mathdutchcal{A}\ui2_g(t)+\mathdutchcal{B}\ui2_g(t)\right)^2\ ,
\end{split}\\\label{eq:g4exp}
\begin{split}
G_4(t) =&\left(1-\frac{t}{4M_N^2}\right) \left(\mathdutchcal{C}_g(t)\right)^2 \ .
\end{split}
\end{align}
We note that these expressions are equivalent to the ones in the previous work~\cite{Guo:2021ibg} if one takes the leading-Mellin-moment approximation. However, since the leading-Mellin-moment approximation is made under the assumption $\xi = 1$, in principle it cannot predict any $\xi$-dependence, i.e., the $\xi$ should be taken to be 1 in this case. The explicit $\xi$-dependence was kept in Ref. \cite{Guo:2021ibg} to partially account for the effect of the relatively small $\xi$ of the measurements. This work, on the other hand, predicts the $\xi$-scaling in the gCFFs and accordingly in the differential cross sections in the large $\xi<1$ region, besides providing a justification for the leading-moment approximation in this region. Therefore, it is crucial to look for such behaviors in the measurements to justify this framework and also to extract these coefficients. 

In this section, we will study the $\xi$-scaling of the differential cross sections of near-threshold $J/\psi$ photo-production utilizing the recently published data from the $J/\psi$ 007 experiment~\cite{Duran:2022xag} as well as the GlueX collaboration~\cite{GlueX:2023pev}. We should also note that, just like in the previous work~\cite{Guo:2021ibg,Guo:2023pqw}, large $\xi$ will still help suppress the systematical uncertainties here. However, since we focus on the $\xi$-scaling of the data itself rather than the actual extraction of gluonic GFFs, we will keep the small-$\xi$ data for comparison.

\subsection{The \texorpdfstring{$\xi$}{XI}-scaling of the \texorpdfstring{$J/\psi$}{jpsi} production data}

\begin{figure}[t]
    \includegraphics[width=0.48\textwidth]{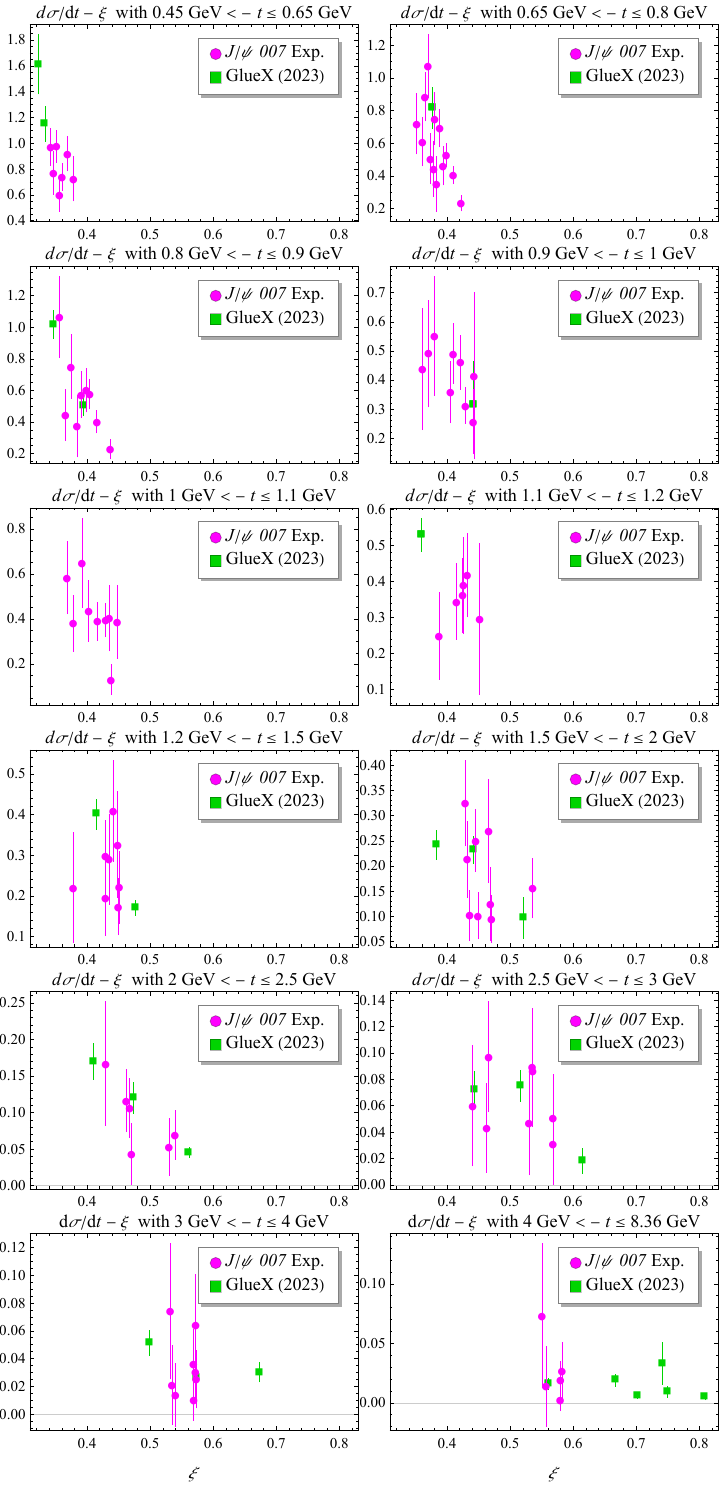}
    \caption
    {\raggedright Differential cross sections $\text{d}\sigma/\text{d}t$ (nb/GeV$^2$) of the threshold $J/\psi$ photo-production versus $\xi$ combing the $J/\psi$ 007 (circle)~\cite{Duran:2022xag} and the GlueX measurements (square)~\cite{GlueX:2023pev}. The measurements are binned into groups with similar $|t|$ and plotted with respect to the $\xi$.}
    \label{fig:difxsecscale}
\end{figure}

To start with, we consider the kinematical coverage of the differential cross-section data. As shown in FIG. \ref{fig:jpsixitplt}, these data roughly cover the near threshold region, although much fewer data exist in the large $|t|$ region with larger uncertainties due to the lack of events.
According to eq. (\ref{eq:xsecscale}), the differential cross sections are expected to have the $\xi^{-4},\xi^{-2}$ and $\xi^0$ scaling behaviors at fixed $|t|$, corresponding to the $G_0(t)$, $G_2(t)$ and $G_4(t)$ at the leading order of the asymptotic expansion. Since we do not have enough measurements at the same $|t|$, we combine the data with relatively close $|t|$ as shown in FIG. \ref{fig:jpsixitplt} with the dashed line. In FIG. \ref{fig:difxsecscale}, we show the $\xi$-dependence of the differential cross-section binning data with similar $|t|$. It is clear from the plots that the differential cross sections have non-trivial dependence on the $\xi$ and generally get suppressed with increasing $\xi$, consistent with the expected $\xi$-scaling behaviors from the asymptotic expansion. However, due to the limited quality of the data overall as well as the kinematical constraint that for given $|t|$ there will be a maximum $\xi$ it could reach:
\begin{equation}
    |t|\ge \frac{4\xi^2}{1-\xi^2} M_{N}^2\ ,
\end{equation}
it is challenging to extract the full $\xi$-dependence at given $|t|$ with the current data. Consequently, one has to consider the combination of all the data with different $|t|$ to study the $\xi$-dependence, for which one has to take the non-trivial $|t|$-dependence into account as well.\footnote{The differential cross section also depends on the beam energy $E_b$ which can be equivalently expressed in terms of the center of mass energy $W$, see e.g., eq. (\ref{eq:xsec}). However, near the threshold $W\sim M_N+M_{J/\psi}$ so the $W$-dependence is weak.}

To test the $\xi$-scaling behavior with all the differential cross-section data of different $|t|$, we consider one of the simplest ansatz for the $\left|G(\xi,t)\right|^2$:
\begin{equation}
\label{eq:gansatz}
    \left|G(\xi,t)\right|^2 = N \xi^{-\alpha} \left(1-\frac{t}{\Lambda^2}\right)^{-6}\ ,
\end{equation}
where $\Lambda$ represents the effective tripole mass (note that $\left|G(\xi,t)\right|^2$ is the square of form factors, so it has a power of $6$ for tripole), $N$ corresponds to the normalization and $\alpha$ indicates the power of the $\xi$-scaling. Fitting to the differential cross-section measurements, we obtain $N=0.027\pm0.007$, $\alpha= 5.17 \pm0.25$ and $\Lambda=3.57\pm0.22$ with reduced $\chi^2=1.23$. The best-fit $\alpha$ is around $5$ which is quite close to the expected value of $4$ from asymptotic expansion. Actually, by fixing $\alpha=4$, we obtain $N=0.096\pm0.005$ and $\Lambda=2.96\pm0.10$ with reduced $\chi^2 = 1.42$. Both fits produce quite reasonable reduced $\chi^2$s, indicating strongly that differential cross sections scale with $\xi$ according to $\xi^{-4}$ or $\xi^{-5}$. 

We note that these fits assume factorizable $t$-dependence as eq. (\ref{eq:gansatz}) which would not be generally true. However, not enough information can be obtained about the potentially entangled $(\xi,t)$-dependence given the present amount of data and factorized $\xi$- and $t$-dependence appears to work fine.

\begin{figure}[t]
    \includegraphics[width=0.48\textwidth]{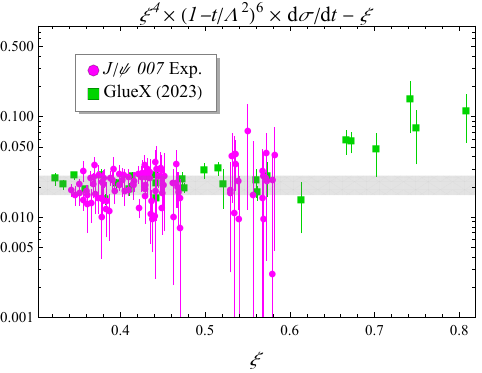}
    \caption
    {\raggedright Rescaled differential cross sections $\xi^4\left(1-t/\Lambda^2\right)^6\text{d}\sigma/\text{d}t$ (nb/GeV$^2$) of the threshold $J/\psi$ photo-production versus $\xi$ combing the $J/\psi$ 007 (circle)~\cite{Duran:2022xag} and the GlueX measurements (square)~\cite{GlueX:2023pev}. The orange band corresponds to the best-fit value and its width is calculated based on the different $E_b$ among all the data (8.7 GeV $<~E_b~<$ 11 GeV).}
    \label{fig:xiscalediffxsec}
\end{figure}

To show the $\xi$-scaling of differential cross sections more clearly, in FIG. \ref{fig:xiscalediffxsec}, we plot the rescaled differential cross sections multiplied to $\xi^4\left(1-t/\Lambda^2\right)^6$. It is apparent that most of the data are statistically consistent with the constant orange band after the rescaling, whereas there appears to be some rising in the $\xi$-dependence according to the GlueX measurements at large $\xi$. This observation has been discussed by the GlueX collaboration that their measured differential cross sections anomalously increase as $|t|$ increases~\cite{GlueX:2023pev}. However, this was not seen in the $J/\psi$ 007 measurements partially due to its limited coverage at large $|t|$. With that in mind, we shall note that this non-trivial behavior could be important for studying the $\xi$-dependence of the differential cross section, although more data are needed to further justify this observation.

At the end of this subsection, we also comment on the remnant $\xi$-dependence besides the overall $\xi^{-4}$-scaling. According to FIG. \ref{fig:xiscalediffxsec}, one might claim that the differential cross sections scale as $\xi^{-4}$ without remnant $\xi$-dependence, and thus the $G_2(t)$ and $G_4(t)$ terms in eq. (\ref{eq:xsecscale}) vanish. However, it is worth noting that the data do not distribute evenly among all the different $\xi$ --- a major part of the data have $0.3<\xi<0.5$ while only about $1/3$ of them have $\xi>0.5$. In addition, these data with large $\xi$ typically have lower quality, and are less weighted statistically. Therefore, the remnant $\xi$-dependence is not excluded by the current data. Attempting to extract the $G_2(t)$ and $G_4(t)$ by fitting to the current data will lead to non-zero best-fit $G_2(t)$ and $G_4(t)$ with massive uncertainties, consistent with the these arguments.

\subsection{Several scenarios for the extraction of gCFFs and gluonic GFFs}

With the above discussion, we explore more the extraction of the gCFFs and even the gluonic GFFs in this subsection, which has been studied in the previous work~\cite{Duran:2022xag,Guo:2021ibg,Guo:2023pqw}. However, as we discussed in the previous subsection, the current differential cross-section data only support the overall $\xi^{-4}$-scaling behavior. Any additional conclusions one attempts to draw from them may suffer from insufficient data, which is consistent with the finding in the previous work that the $C_g(t)$ form factors cannot be effectively constrained with the current data~\cite{Duran:2022xag,Guo:2021ibg,Guo:2023pqw}. Therefore,
we note that it still crucial to obtain more data with higher quality, especially at large momentum transfer $|t|$, to get further information on the gCFFs or even the gluonic GFFs. However, to illustrate how the analysis could be done, we will use the current data as an example to consider several scenarios that the gCFFs or gluonic GFFs can be extracted.

\subsubsection*{Scenario 1: Leading-moment approximation}

The simplest way to obtain the gluonic GFFs is through the leading-moment approximation, with either the leading-Mellin-moment approximation or the leading-conformal-moment approximation that differs by a factor of $\mathcal{A}_1^{\rm{conf}}=5/4$. The analysis reduces to the ones in the previous work~\cite{Duran:2022xag,Guo:2021ibg,Guo:2023pqw} in the leading-Mellin-moment approximation. Extensive analyses of the current data with careful treatments of the leading-moment approximation and large-$\xi$ expansion have been presented in Ref. \cite{Guo:2023pqw} recently.

Here we make two more remarks: first, in Ref. \cite{Guo:2023pqw} the leading-Mellin-moment approximation has been applied to the data with $\xi>0.4$ due to the lack of large-$\xi$ data. However, the example in FIG. \ref{fig:xiscale} indicates that the leading-moment approximation works best with $\xi>0.6$ while there are noticeable interferences from the $\xi^{-4}$ term when extending to $\xi>0.3$. The corresponding systematic uncertainties from the $\xi^{-4}$ term can be roughly estimated to be $20-30\%$ based on the difference between eqs. (\ref{eq:xi2termfit}) and (\ref{eq:xi3termfit}). Second, the extra factor of $\mathcal{A}_1^{\rm{conf}}=5/4$ will lead to a slightly different result for the extraction of gluonic GFFs, which agrees better to the gluonic GFFs from lattice simulation~\cite{Pefkou:2021fni}.

\subsubsection*{Scenario 2: Modified leading-moment approximation}

Motivated by the above observations, a modified approach can be proposed by multiplying an extra normalization factor to the leading-moment approximation. There are several benefits to doing so. First, this extra normalization will take care of the difference between the leading-Mellin-moment and leading-conformal-moment approximation as well as other potential higher-moment effects. Second, the numerical example in FIG. \ref{fig:xiscale} indicates that the extracted coefficients may be off by an extra factor when including lower-$\xi$ data. This could be partially accounted for by the extra normalization. Third, there might be overall normalization from the higher-order effects that could be absorbed into the extra normalization. Therefore, an extra normalization seems reasonable, especially when lower-$\xi$ data will be included, although it cannot represent all the other effects. Thus, we consider the approximation
\begin{eqnarray}\label{eq:modleadingConfMomA}
    \mathdutchcal{A}_g\ui{2}(t)&\approx 2 N_{\rm{gCFF}}~\mathcal{A}_1^{\rm{conf}} A_g(t)\ ,\\\label{eq:modleadingConfMomC}
    \mathdutchcal{C}_g(t)&\approx 8 N_{\rm{gCFF}} ~\mathcal{A}_1^{\rm{conf}} C_g(t)\ .
\end{eqnarray}
with $N_{\rm{gCFF}}$ the extra normalization constant to be determined, where the $\mathdutchcal{B}_g\ui{2}(t)$ will still be set to 0.

However, with the current data even including the lower-$\xi$ ones, the extraction of the gCFFs with the extra normalization $N_{\rm{gCFF}}$ still suffers from insufficient data, especially for the $\mathdutchcal{C}_g(t)$ coefficient. We remark that it really takes the non-trivial $\xi$-dependence to separate the two gluonic GFFs. This will be discussed with more details in the next section.

\subsubsection*{Scenario 3: Extraction with complete \texorpdfstring{$\xi$}{XI}-dependence}
In the idea case where the amount and quality of the data are unlimited, we could consider the most model-independent way to extract the gCFFs and gluonic GFFs. In this case, we would first obtain the coefficients $G_0(t)$, $G_2(t)$, and $G_4(t)$ in eqs. (\ref{eq:g0exp})--(\ref{eq:g4exp}) from the full $\xi$-dependence of the differential cross sections. Then we could try to reconstruct the coefficients $\mathdutchcal{A}_g\ui{2}(t)$, $\mathdutchcal{B}_g\ui{2}(t)$, and $\mathdutchcal{C}_g(t)$ or even the leading moments $A_g(t)$, $B_g(t)$, and $C_g(t)$ from them. Even though the reconstruction in the second step still suffers from the interference of the higher moments, the extraction of $G_0(t)$, $G_2(t)$, and $G_4(t)$ in the first step could avoid the potential contamination from the higher-order terms or the imaginary part of the gCFFs that scale differently in $\xi$. More importantly, such extraction allows a direct test of the factorization and the asymptotic expansion with gluon GPDs, i.e., the differential cross sections should have the $\xi$-scaling behaviors as predicted by eq. (\ref{eq:xsecscale}) to the leading order independently of the specific values of the gluon GPDs.

In the previous subsection, we showed that the overall $\xi$-scaling of the differential cross section is consistent with the expected $\xi^{-4}$ behavior whereas no further information can be unambiguously obtained. Although the GlueX measurements seem to indicate some non-trivial remnant $\xi$-dependence in the differential cross section as $\xi$ get large, we note that such observations also depend on how one parameterize the $t$-dependence, and it is crucial to have more data to confirm such observations. Moreover, we note that this scenario could apply to a larger range of $\xi$ as long as the asymptotic $\xi$-scaling behavior is satisfied.

\section{Current status and future developments}
\label{sec:currentstatus}
The above scenarios certainly put non-trivial requirements on both the quantity and the quality of the available data, which cannot be fulfilled with the current $J/\psi$ threshold production measurements, especially in the large-$|t|$ region. Accordingly, in this section we will discuss the impact of the current data as well as the future developments on the extraction of the gluonic GFFs in the GPD factorization framework. We will focus on the current limitations in the extraction and the possible improvements from future developments.

\subsection{Current status of the gluonic GFFs extraction}

To start with, we shortly summary the current status of the extraction of the gluonic GFFs in the GPD framework. These extractions are all under the scenario 1 where the leading-moment approximation is made due to the limited amount of data~\cite{Duran:2022xag,Guo:2021ibg,Guo:2023pqw}. The most up-to-date analyses have been presented in Ref.~\cite{Guo:2023pqw}. As discussed therein, the gluonic GFFs still cannot be fully determined from the current $J/\psi$ threshold production measurements alone even under the leading-moment approximation. Typically, the $A_g(t)$ and $C_g(t)$ form factors are parameterized in the tripole form, and then the $A_g(0)$ form factor is fixed by the gluon PDF in the forward limit that $A_g(0)= 0.414$ to make the extraction of GFFs feasible~\cite{Hou:2019efy}.
With such a setup, the $A_g(t)$ form factor could be reasonable constrained/ extracted from the $J/\psi$ threshold production measurements, whereas the $C_g(t)$ form factor still could not.

It is noteworthy that the squared hadronic matrix element $\left|G(\xi,t)\right|^2$ reads,
\begin{align}\label{eq:Gscaling2}
\begin{split}
    \left|G(\xi,t)\right|^2\approx\xi^{-4} \Big[&\left(1-\xi^2\right)\left(\mathdutchcal{A}_g\ui{2}(t)\right)^2+2 \mathdutchcal{A}_g\ui{2}(t)  \xi^{2} \mathdutchcal{C}_g(t)\\
    &+\left(1-\frac{t}{4M_N^2}\right) \left(\xi^2\mathdutchcal{C}_g(t)\right)^2\Big]+\cdots\ ,
\end{split}
\end{align}
when ignoring $\mathdutchcal{B}_g\ui{2}(t)$. Based on the gluon PDF, we have $\mathdutchcal{A}_g\ui{2}(0) \approx 5/2 A_g(0)\approx 1$~\cite{Hou:2019efy}, and we also estimate $\mathdutchcal{C}_g(0) \approx 10 C_g(0)\approx -5$ from lattice QCD simulations~\cite{Pefkou:2021fni}. Therefore, for the major part of the current data with $0.3<\xi<0.5$, we have roughly $0.45<\xi^2\left|\mathdutchcal{C}_g(0)\right|<1.25$ which is comparable to $\mathdutchcal{A}_g\ui{2}(0)\approx 1$. Thus, we have $\mathdutchcal{A}_g\ui{2}(0) \sim \xi^2\left|\mathdutchcal{C}_g(0)\right|$, indicating the $A_g(t)$ and $C_g(t)$ form factors could have comparable contributions to the squared hadronic matrix element, and the sensitivities of the data to them could be comparable accordingly. Moreover, the sensitivity to the $A_g(t)$ form factor will be suppressed with increasing $\xi$ by the prefactor of $(1-\xi^2)$. Thus, one should expect increasing sensitivity to the $C_g(t)$ form factors when large-$\xi$ data are obtained. We note that this picture will be modified when considering their generally different $t$-dependence and the large systematical uncertainties associated with the leading-moment approximation in the lower-$\xi$ region. 

However, the above estimation is rather contrary to the observation in the current extraction that the $A_g(t)$ form  factor is much better constrained~\cite{Duran:2022xag,Guo:2021ibg,Guo:2023pqw}, mainly due to the extra off-forward constraint on the $A_g(t)$ form  factor from gluon PDFs. Therefore,  besides the systematical uncertainties, one should keep such model-dependence in mind when extracting the gluonic GFFs with leading-moment approximation and fixed $A_g(0)$. More coverage of large-$\xi$ region will be crucial to gain more sensitivity of the $C_g(t)$ form factor and to clarify such model-dependence. Actually, the result in FIG. \ref{fig:xiscalediffxsec} clearly indicates that only one coefficient is effectively constrained. Accordingly, extraction of any extra information requires extra input, as what has been done in Refs. \cite{Duran:2022xag,Guo:2021ibg,Guo:2023pqw}.

To illustrate the effect of the large-$\xi$ data on the extraction of the GFFs, particularly the $C_g(t)$ form factor, we consider the different $\xi$-scaling behaviors corresponding to different values of the $C_g(t)$ form factor. From eq. (\ref{eq:Gscaling2}) it is clear that the $\xi$-scaling of the differential cross sections depends on the ratio $r_C(t)\equiv \mathdutchcal{C}_g(t)/\mathdutchcal{A}_g\ui{2}(t)$ which corresponds to $4 C_g(t)/A_g(t)$ or $D_g(t)/A_g(t)$ in terms of the gluonic GFFs in the leading-moment approximation. Generally, the ratio $r_C(t)$ will have residual $t$-dependence, which will be neglected for discussions. We consider three cases where $r_C(t)\sim +1$, $r_C(t)\sim -1$ and $r_C(t)\sim 0$.\footnote{Based on the lattice simulation~\cite{Pefkou:2021fni}, $r_C(t)$ ranges from around $-1$ to $-5$ with $\mathcal{O}(1)$ uncertainties when $t$ varies from about $-0.1$ GeV$^2$ to $-2$ GeV$^2$.} Then the corresponding $\xi$-scaling behaviors of the $\left|G(\xi,t)\right|^2$ will be:
\begin{eqnarray}
    \left|G(\xi,t)\right|^2&\sim&\xi^{-4}\left(1-\xi^2\right), \qquad\qquad~ r_C(t) \sim 0\\
    \left|G(\xi,t)\right|^2&\sim&\xi^{-4}\left(1-3\xi^2+2\xi^4\right), \quad~  r_C(t)  \sim -1\\
    \left|G(\xi,t)\right|^2&\sim&\xi^{-4}\left(1+\xi^2+2\xi^4\right),\qquad r_C(t) \sim +1
\end{eqnarray}
respectively, where we assume the overall tripole $t$-dependence and approximate $-t/(4M_N^2) \sim 1$ for simplicity. Then we use similar ansatzes to eq. (\ref{eq:gansatz}) to examine the different $\xi$-scaling behaviors. For comparison, we fix the tripole mass $\Lambda = 3$ GeV from the previous fit and only fit the different normalization prefactor $N$ to the data for each of them.

\begin{figure}[t]
    \includegraphics[width=0.48\textwidth]{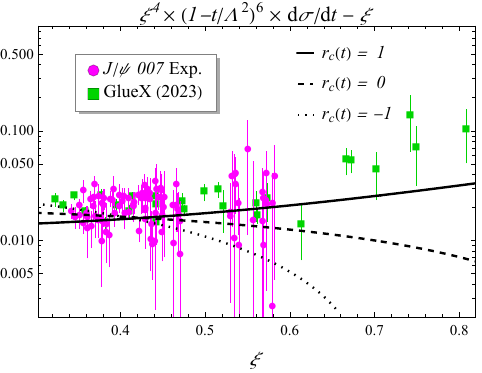}
    \caption
    {\raggedright Rescaled differential cross sections $\xi^4\left(1-t/\Lambda^2\right)^6\text{d}\sigma/\text{d}t$ (nb/GeV$^2$) with different $r_C(t)$ of which the normalization factors are fixed by fitting to all the differential cross-section data. The solid, dashed, and dotted lines correspond to fits with $r_C=1$, $r_C=0$, and $r_C=-1$ with fixed $E_b=11$ GeV. Note that the differential cross section depends on $E_b$ but only weakly.}
    \label{fig:xiscalediffxseccomp}
\end{figure}

\begin{figure}[t]
    \includegraphics[width=0.48\textwidth]{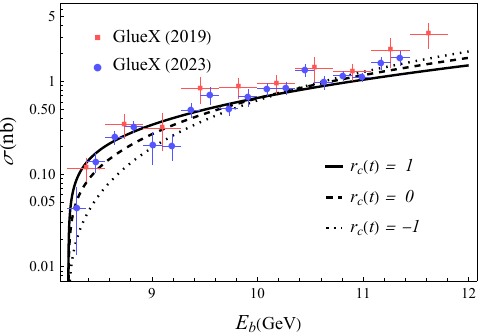}
    \caption
    {\raggedright Total cross sections $\sigma(\text{nb})$ with different $r_C(t)$ compared to the GlueX measurements. The solid, dashed, and dotted lines correspond to fits with $r_C=1$, $r_C=0$, and $r_C=-1$. Normalization factors are determined by the differential cross section only.}
    \label{fig:sigmarplts}
\end{figure}
In FIG. \ref{fig:xiscalediffxseccomp}, we show the three rescaled differential cross sections $\xi^4\left(1-t/\Lambda^2\right)^6\text{d}\sigma/\text{d}t$ with different $r_C(t)$. In FIG. \ref{fig:sigmarplts}, we show the direct comparisons of the total cross section with different $r_C(t)$. All normalization factors are determined by fitting to all differential cross-section data, and all three fits have reasonable reduced $\chi^2$s: $1.34$ for $r_C(t) = 0$, $1.64$ for $r_C(t) = -1$ and $1.66$ for $r_C(t) =1$. Besides, we have several remarks regarding these results.
First, we emphasize that these fits should not be regarded as extractions of gluonic GFFs since the actually $t$-dependence could be much more complicated. Moreover, these different sets show clearly that the data do not uniquely determine the remnant $\xi$-scaling. For instance, in FIG. \ref{fig:xiscalediffxseccomp} the rising solid line corresponding to $r_C=1$ appears to agree better with the data overall, which is NOT supported by the reduced $\chi^2$ which is dominated by small/medium-$\xi$ data. Thus, more coverage in the large-$\xi$ region and higher quality of the data are crucial for further justification. As mentioned before, there are two main corrections due to the higher moments and the non-vanishing imaginary part. The higher moments will modify the coefficients of lower power of $\xi$, namely the $G_0(t)$, $G_2(t)$ and $G_4(t)$ and also cause higher power of $\xi$, namely $G_6(t)$, $G_8(t)$ and even higher. Explicitly including these higher moments would improve the extraction, which is, nevertheless, impractical given the difficulty in obtaining them. Thus, we note that there will be corrections from the higher moments, though they are generally suppressed compared to the leading ones and can be partially separated with their different $\xi$-scaling. As for the non-vanishing imaginary part, its behavior at lower $\xi$ is generally not known, which could cause significant systematic uncertainties and break the $\xi$-scaling behavior in the lower-$\xi$ region. Thus, the usage of the lower-$\xi$ data for this analysis should be avoided if possible unless the impact of the imaginary CFF is properly considered and extra care should be taken in the lower-$\xi$ region.

\subsection{Impact of the future facilities and experiments}

To obtain better extractions beyond the above results, more experimental inputs are necessary. Here we discuss several important future developments and their impact on the study of the gluonic structures with threshold heavy meson production in the GPD framework.

The planned Solenoidal Large Intensity Device (SoLID) detector for the Hall A at JLab is a large acceptance spectrometer capable of handling high luminosity ~\cite{JeffersonLabSoLID:2022iod}. This would allow an unprecedented precision measurement of the differential cross sections $J/\psi$ production near the threshold over a large kinematical range. As we discussed in the previous section, both large kinematical coverage and high precision are crucial for the study of the $\xi$-scaling behaviors and the potential extraction of the gCFFs and gluonic GFFs. This would be much easier with the future SoLID detector.

Moreover, there has been rising interest in pursuing an energy upgrade to 20 GeV or higher of the Continuous Electron Beam Accelerator Facility (CEBAF) at JLab~\cite{Accardi:2023chb}. For the threshold heavy quarkonium photo-production, this energy upgrade will bring in the new possibility for measuring the production of $\psi'$ or $\psi(2S)$~\cite{JeffersonLabSoLID:2022iod}, the first excited state of $J/\psi$. Its slightly heavier mass $M_{\psi'}=3.686$ GeV will be beneficial in suppressing the higher-order effects as well as approaching the larger-$\xi$ region. Although the improvement might not be significant given the similar masses of $J/\psi$ and $\psi'$, the simultaneous measurement of the productions of two distinguishable but similar particles would allow a direct examination of the production mechanism, i.e., the factorization and the universality of the GPD.

Looking into the further future, we also have the EIC~\cite{AbdulKhalek:2021gbh,Burkert:2022hjz}, of which the much higher center-of-mass energy will provide a unique opportunity for studying the production of the heavier quarkonium like the $\Upsilon$ with quasi-real photon~\cite{Gryniuk:2020mlh}. It will be harder to measure the production close to the threshold with colliders, of which simulations show that one can get the center-of-mass energy $W \gtrsim 12$ GeV making use of the low-energy setting of EIC, noting that $W_{\rm{th}}=10.4$ GeV. Consequently, approaching the large-$\xi$ region near the threshold with $\xi_{\rm{th}}\sim 0.8$ for $\Upsilon$ production could be challenging for EIC. However, we still have $\xi\gtrsim 0.5$ at $W\sim 12$ GeV for $\Upsilon$ production, improving from the $\xi \gtrsim 0.35$ in the case of $J/\psi$ production. Moreover, the much larger mass $M_{\Upsilon} = 9.46$ GeV will be extremely helpful in suppressing the higher-order corrections, which have not been systematically studied yet.

Besides, we note that the polarized measurements will be extremely helpful as well. Since different target polarizations correspond to different combinations of the gCFFs in the amplitude, they could serve as a direct examination of the production mechanism as well, similar to the $\psi'$ or $\Upsilon$ productions. Moreover, with different target polarizations, the sensitivity to different combinations of the gCFFs will facilitate the disentanglement of these gCFFs, and enhance the extraction eventually.

\section{Conclusion and comments}
\label{sec:conc}
To conclude, we study the exclusive productions of heavy quarkonium near the threshold utilizing the large-skewness behaviors of GPDs. The endpoint constraints at $|x|=1$ suppress the PDF-like region $\xi<|x|<1$ in the large-$\xi$ limit, and cause GPDs' behaviors to be dominated by the DA-like regions. Consequently, the Compton-like amplitudes can be written as an asymptotic series in terms of the moments of GPDs. We examined such possibilities with a double distribution parameterization of GPDs and another parameterization based on conformal moments. We find that in the large-$\xi$ limit the real parts of the Compton-like amplitudes in these cases are indeed well approximated by the asymptotic series with superasymptotic approximation, whereas the imaginary parts are suppressed.

We then apply the above observation to the recent measurements of the threshold $J/\psi$ production by the $J/\psi$ 007 experiment~\cite{Duran:2022xag} and the GlueX collaboration~\cite{GlueX:2023pev}. We find that the $\xi$-scaling of the measured differential cross sections is consistent with the asymptotic predictions, although the extraction of further information is limited by the quality and the $\xi$-coverage of the current data. More specifically, while the measured $t$-dependence of the differential cross sections could effectively constrain the overall $t$-dependence of the form factors, more coverage in the large $\xi$ region is crucial to separate the contributions of the $A_g(t)$ and $C_g(t)$ form factors, assuming that $A_g(0)$ can be obtained from forward gluon PDFs. We also present several scenarios for the extraction of the gluonic GFFs from such processes and discuss the impact of the future experimental developments.

We would like to emphasize that our analyses in this paper were based on the leading order perturbative calculations of the heavy-quarkonium production in the GPD formalism. We expect the generic features of the above results will remain even at higher order approximation in strong coupling. Of course, it will be highly desired to pursue such analysis at the next-to-leading order (NLO). We notice that the NLO calculations for exclusive photo-production of heavy quarkonium states have been carried out in the literature for high-energy scattering process, i.e., at small skewness~\cite{Ivanov:2004vd,Chen:2019uit, Flett:2021ghh}. When these calculations are extended to more general kinematics, including near threshold region, we can further check the asymptotic expansion relations found in our paper.

Furthermore, we note that since the analyses in this work mostly rely on the large-$\xi$ kinematics and the endpoint behaviors of the GPDs, it may be possible to implement them to the quark GPDs with similar processes like the photo-production of lepton pairs. The kinematical setup overlaps with that of the time-like Compton scattering (TCS)~\cite{Berger:2001xd,CLAS:2021lky} but large skewness will be preferred. There has been work on the extraction of the quark GFFs $C_q(t)$ with deeply virtual Compton scattering (DVCS)~\cite{Burkert:2018bqq,Kumericki:2019ddg}. Although they are under a different framework with the dispersion relation using small/medium-$\xi$ data, this work, especially the discussion at the end of the sec. \ref{sec:asymptotic} on the connection to the dispersion relation, seems to provide a justification of the analyses therein from the large-$\xi$ perspective. Such sensitivities to the $C_q(t)$ form factor (which is essentially the $\mathdutchcal{C}_q(t)$ coefficient analogous to the gluonic one $\mathdutchcal{C}_g(t)$) are encouraging for future studies.

\begin{acknowledgments}
We thank Yoshitaka Hatta for useful discussions and correspondences. This material is based upon work supported by the LDRD program of LBNL, and by the U.S. Department of Energy, Office of Science, Office of Nuclear Physics, under contract numbers DE-AC02-05CH11231. The authors also acknowledge partial support by the U.S. Department of Energy, Office of Science, Office of Nuclear Physics under the umbrella of the Quark-Gluon Tomography (QGT) Topical Collaboration with Award DE-SC0023646.
\end{acknowledgments}

\bibliographystyle{apsrev4-1}
\bibliography{refs.bib}
\end{document}